%Paper: hep-ph/9304277
%From: My Account <me@cryptons.tamu.edu>
%Date: Tue, 20 Apr 93 12:17:00 -0800

%
% The three figures are available upon request from me@cryptons.tamu.edu
% as one uuencoded file or three separate postscript files. The uuencoded file
% is the preferred option.
\input harvmac

\def\Titleh#1#2{\nopagenumbers\abstractfont\hsize=\hstitle\rightline{#1}%
\vskip .5in\centerline{\titlefont #2}\abstractfont\vskip .5in\pageno=0}

\def\CTPa{\it Center for Theoretical Physics, Department of Physics,
      Texas A\&M University}
\def\CTPb{\it College Station, TX 77843-4242, USA}
\def\HARCa{\it Astroparticle Physics Group,
Houston Advanced Research Center (HARC)}
\def\HARCb{\it The Woodlands, TX 77381, USA}

\def\ie{\hbox{\it i.e.}}     
\def\eg{\hbox{\it e.g.}}

\def\coeff#1#2{{\textstyle{#1\over #2}}}

\catcode`\@=11 % This allows us to modify PLAIN macros.

\def\lsim{\mathrel{\mathpalette\@versim<}}
\def\gsim{\mathrel{\mathpalette\@versim>}}
\def\@versim#1#2{\vcenter{\offinterlineskip
    \ialign{$\m@th#1\hfil##\hfil$\crcr#2\crcr\sim\crcr } }}
\def\boxit#1{\vbox{\hrule\hbox{\vrule\kern3pt
      \vbox{\kern3pt#1\kern3pt}\kern3pt\vrule}\hrule}}

\def\etal{{\it et. al.}}

\def\t1{{\tilde 1}}

\def\JL{J. L. Lopez}
\def\DVN{D. V. Nanopoulos}
\def\AZ{A. Zichichi}
\def\HP{H. Pois}
\def\XW{X. Wang}

\def\GeV{\,{\rm GeV}}

\def\bsg{b\to s\gamma}
\def\brbsg{{\rm BR}(\bsg)}
\def\NPB#1#2#3{Nucl. Phys. B {\bf#1} (19#2) #3}
\def\PLB#1#2#3{Phys. Lett. B {\bf#1} (19#2) #3}

\def\PRD#1#2#3{Phys. Rev. D {\bf#1} (19#2) #3}
\def\PRL#1#2#3{Phys. Rev. Lett. {\bf#1} (19#2) #3}
\def\PRT#1#2#3{Phys. Rep. {\bf#1} (19#2) #3}

\def\TAMU#1{Texas A \& M University preprint CTP-TAMU-#1}

\nref\CLEO{M. Battle, \etal\ (CLEO Collab.) in Proceedings of the joint
Lepton-Photon and Europhysics Conference on High-Energy Physics, Geneva
1991.}
\nref\Barger{V. Barger, M. Berger, and R. J. N. Phillips, \PRL{70}{93}{1368}.}
\nref\Hewett{J. Hewett, \PRL{70}{93}{1045}.}
\nref\newCLEO{E. Thorndike, CLEO Collab., talk given at the 1993 Meeting of
the American Physical Society, Washington D.C., April 1993.}
\nref\VernonHARC{V. Barger, talk presented at the HARC workshop ``Recent
Advances in the Superworld", The Woodlands, Texas, April 1993.}
\nref\Bertolini{S. Bertolini, F. Borzumati, A. Masiero, and G. Ridolfi,
\NPB{353}{91}{591}.}
\nref\BG{R. Barbieri and G. Giudice, CERN-TH.6830/93.}
\nref\Dickreview{For reviews see: R. Arnowitt and P. Nath, {\it Applied N=1
Supergravity} (World Scientific, Singapore 1983);
H. P. Nilles, \PRT{110}{84}{1}.}
\nref\LNZb{\JL, \DVN, and \AZ, \TAMU{68/92}, CERN-TH.6667/92, and
CERN-PPE/92-188.}
\nref\GSW{See \eg, B. Grinstein, R. Springer, and M. Wise, \NPB{339}{90}{269};
M. Misiak, \PLB{269}{91}{161}, and references therein.}
\nref\AG{A. Ali and C. Greub, \PLB{293}{92}{226}.}
\nref\MisiakII{M. Misiak, ZH-TH-19/22 (1992).}
\nref\aspects{S. Kelley, \JL, \DVN, H. Pois, and K. Yuan, \TAMU{16/92} and
CERN-TH.6498/92 (to appear in Nucl. Phys. B).}
\nref\AN{R. Arnowitt and P. Nath, \PRL{69}{92}{725}; P. Nath and R.
Arnowitt, \PLB{287}{92}{89} and \PLB{289}{92}{368}.}
\nref\LNZa{\JL, \DVN, and \AZ, \PLB{291}{92}{255}.}
\nref\LNP{\JL, \DVN, and \HP, \PRD{47}{93}{2468}.}
\nref\ANcosm{R. Arnowitt and P. Nath, \PLB{299}{93}{58} and Erratum;
P. Nath and R. Arnowitt, NUB-TH-3056/92, CTP-TAMU-66/92 (revised).}
\nref\LNPZ{\JL, \DVN, \HP, and \AZ, \PLB{299}{93}{262}.}
\nref\LN{For a review see, A. B. Lahanas and D. V. Nanopoulos,
\PRT{145}{87}{1}.}
\nref\LNPWZh{\JL, \DVN, \HP, \XW, and \AZ,  CERN-PPE/93-17 and \TAMU{05/93}
(to appear in Phys. Lett. B).}
\nref\LNWZ{\JL, \DVN, \XW, and \AZ, CERN-PPE/92-194, CTP-TAMU-76/92.}
\nref\LNPWZ{\JL, \DVN, \HP, \XW, and \AZ, \TAMU{89/92}, CERN/LAA/93-01,
CERN-TH.6773/93, and CERN-PPE/93-16.}
\nref\hera{\JL, \DVN, \XW, and \AZ, CTP-TAMU-15/93, CERN/LAA/93-19.}
\nref\nino{A. Zichichi, \etal, in preparation.}

\nfig\I{The calculated values of $\brbsg$ versus the chargino mass (top row) in
the minimal $SU(5)$ supergravity model. The horizontal dashed line represents
the new CLEO upper bound. Also shown (bottom row) are the ratios
of these branching fractions to the corresponding SM values.}
\nfig\II{The calculated values of $\brbsg$ versus the chargino mass (top
row) in the no-scale flipped $SU(5)$ supergravity model. Note the fraction of
parameter space excluded by the new CLEO upper bound, denoted by the horizontal
dashed line. Also shown (bottom row) are the ratios of these branching
fractions to the corresponding SM values plotted against the charged Higgs
mass. Note the large suppression of $\brbsg$ that occurs for $\mu>0$ for a good
portion of the parameter space. In this Figure $m_t=100,130,160\GeV$.}
\nfig\III{Same as Fig. 2 but for the strict no-scale scenario. The values of
$m_t$ used are as indicated.}

%\special{landscape}
%\baselineskip=20pt plus 2pt minus 2pt %double space for PRD
\Titleh{\vbox{\baselineskip12pt
\hbox{CTP--TAMU--16/93}
\hbox{ACT--06/93}}}
{\vbox{\centerline{Probing Supergravity Models with}
\centerline{the $\bsg$ Microscope}}}
\centerline{JORGE~L.~LOPEZ, D.~V.~NANOPOULOS, and GYE~T.~PARK,}
\smallskip
\centerline{\CTPa}
\centerline{\CTPb}
\centerline{and}
\centerline{\HARCa}
\centerline{\HARCb}
\vskip .5in
\centerline{ABSTRACT}
We present a calculation of the branching ratio $\brbsg$ in two well motivated
supersymmetric models: the minimal $SU(5)$ and the no-scale flipped $SU(5)$
supergravity models. We find that the improved CLEO upper bound
($\brbsg<5.4\times10^{-4}$ at 95\% CL) does not yet constrain the minimal
$SU(5)$ supergravity model, where $\brbsg_{minimal}=(2.3-3.6)\times10^{-4}$.
In the flipped $SU(5)$ model the CLEO bound is constraining, although still
not very significantly, even for light charged Higgs masses. An improvement in
sensitivity by a factor of two will probe all (more than half) of the parameter
space of the minimal (flipped) $SU(5)$ supergravity model. This ``resolution"
of the $\bsg$ microscope far surpasses that of present collider experiments.
In the flipped model there exists a significant region of parameter space
where $\brbsg$ is highly suppressed due to a new phenomenon involving a
complicated cancellation against the QCD correction parameter.

\bigskip
\Date{April, 1993}

There has recently been a renewed surge of interest on the
flavor-changing-neutral-current (FCNC) $\bsg$ decay, prompted by the
CLEO bound $\brbsg<8.4\times10^{-4}$ at $90\%$ CL \CLEO. Since the Standard
Model (SM) prediction looms around $(2-5)\times10^{-4}$ depending on the
top-quark mass ($m_t$), a reappraisal of beyond the SM contributions has become
topical. Most recently it has been pointed out \refs{\Barger,\Hewett} that the
charged Higgs ($H^\pm$) contributions in two-Higgs-doublet models (like the
minimal supersymmetric standard model (MSSM)) would violate the CLEO bound if
$m_{H^\pm}$ is too light. The precisely excluded region is $m_t$- and
$\tan\beta$-dependent and appears to depend non-negligibly on the bottom-quark
mass ($m_b$) which partially determines the QCD enhancements to the $\bsg$
amplitude. The recently announced improved CLEO bound of
$\brbsg<5.4\times10^{-4}$ at 95\% CL \newCLEO, leads to yet larger excluded
domains in the charged Higgs parameter space \VernonHARC. In these papers it
remained an open question whether inclusion of the full set of supersymmetric
contributions to the $\bsg$ amplitude would change significantly the quoted
bounds, and thus the outlook for supersymmetric Higgs searches at LEPII. The
full supersymmetric contributions to $\bsg$ have been known for some time
\Bertolini\ and have been recently given in a justifiably simplified form in
Ref. \BG. The latter paper argues that the bound on $m_{H^\pm}$ in Refs.
\refs{\Barger,\Hewett} could well be evaded in the full supersymmetric
calculation since in the unbroken supersymmetric limit the $\bsg$ amplitude
vanishes.

The purpose of this paper is to study $\brbsg$ in two supergravity models:
(i) the minimal $SU(5)$ supergravity model \Dickreview, and (ii) the no-scale
flipped $SU(5)$ supergravity model \LNZb. We show that the {\it new} CLEO
bound does not yet constrain the minimal $SU(5)$ model, while some as-yet-mild
constraints are imposed on the flipped $SU(5)$ model, where a new phenomenon
can drastically suppress the $\bsg$ amplitude. We present the results for
$\brbsg$ in these two models and show that improved sensitivity could probe
them in ways not possible at present collider experiments.

We use the following expression for the branching ratio $\bsg$ \BG
\eqn\A{{\brbsg\over{\rm BR}(b\to ce\bar\nu)}={6\alpha\over\pi}
{\left[\eta^{16/23}A_\gamma
+\coeff{8}{3}(\eta^{14/23}-\eta^{16/23})A_g+C\right]^2\over
I(m_c/m_b)\left[1-\coeff{2}{3\pi}\alpha_s(m_b)f(m_c/m_b)\right]},}
where $\eta=\alpha_s(M_Z)/\alpha_s(m_b)$, $I$ is the phase-space factor
$I(x)=1-8x^2+8x^6-x^8-24x^4\ln x$, and $f(m_c/m_b)=2.41$ the QCD
correction factor for the semileptonic decay. The $A_\gamma,A_g$ are the
coefficients of the effective $bs\gamma$ and $bsg$ penguin operators
evaluated at the scale $M_Z$. Their simplified expressions are given in
the Appendix of Ref. \BG, where the gluino and neutralino contributions have
been justifiably neglected \Bertolini\ and the squarks are considered
degenerate in mass, except for the $\tilde t_{1,2}$ which are significantly
split by $m_t$. This is a fairly good approximation to the actual result
obtained in the two supergravity models we consider below, since $m_{\tilde
q}>200\GeV$ in these  models. To include at least partially the QED corrections
to the $\bsg$ operator, in Eq. \A\ we take $\alpha=\alpha(m_b)=1/131.2$.
We use the 3-loop expressions for $\alpha_s$ and choose $\Lambda_{QCD}$ to
obtain $\alpha_s(M_Z)$ consistent with the recent measurements at LEP.
For the numerical evaluation of Eq. (1) we take $\alpha_s(M_Z)=0.118$,
${\rm BR}(b\to ce\bar\nu)=10.7\%$, $m_b=4.8\GeV$ and $m_c/m_b=0.3$.

The subject of the QCD corrections to $\brbsg$, \ie, the origin of the $\eta$
factors and the $C$-coefficient in Eq. \A, has received a great deal of
attention over the years in the SM. The results for supersymmetric models
are not known, but are not expected to differ significantly from the SM
result in the models we consider, since all strongly interacting sparticle
masses are above $M_Z$. In our analysis we use the leading-order QCD
corrections to the $\bsg$ amplitude when evaluated at the $\mu=m_b$ scale
\GSW, \ie, $C=\sum_{i=1}^8 b_i\eta^{d_i}=-0.1766$ for $\eta=0.548$, with the
$b_i,d_i$ coefficients given in Ref. \BG.  The result follows from the
renormalization-group scaling from the scale $M_Z$ down to $\mu=m_b$ of the
effective $\bsg$ operators at $M_Z$, which include the usual electromagnetic
penguin operator plus some four-quark operators. Scaling introduces operator
mixing effects (as exemplified by the appearance of the gluonic penguin
operator in Eq. \A) and the scaled coefficients at $\mu=m_b$ are given by
linear combinations of the coefficients at scale $M_Z$. It has been pointed out
\AG\ that the low-energy mass scale $\mu$ affects the leading-order results
significantly.\foot{This phenomenon is evidenced by the discrepancies in the
bounds on $m_{H^\pm}$ in Ref. \Barger\ and \Hewett, which used
$\mu=m_b=4.25\GeV$ and $\mu=m_b=5.0\GeV$ respectively.} The combined effect on
$\brbsg$ of the uncertainties in $\mu$ and $\Lambda_{QCD}$ has been estimated
to be $\lsim25\%$ \BG. Recently a partial next-to-leading order
calculation of the QCD effects has appeared \MisiakII, which gives somewhat
smaller enhancement factors than the complete leading-order result. A full
next-to-leading order QCD calculation should decrease the above mentioned
uncertainties significantly.

The two models we consider are built within the framework of supergravity with
universal soft-supersymmetry breaking. The renormalization-group scaling
from the unification scale down to low energies plus the requirement of
radiative electroweak symmetry breaking using the one-loop effective potential,
reduces the number of parameters needed to describe these models down to
just five: $m_t$, $\tan\beta$, and three soft-supersymmetry breaking parameters
($m_{1/2},m_0,A$) (for a detailed account of this procedure see \eg, Ref.
\aspects). The sign of the superpotential Higgs mixing term $\mu$ remains as a
discrete variable. The models we consider belong to this general
class of models but are further constrained making them quite predictive.

The minimal $SU(5)$ supergravity model \Dickreview\ is strongly constrained by
the proton lifetime \AN\ and the cosmological constraint of a not too young
universe \refs{\LNZa,\LNP,\ANcosm}. A thorough exploration of the parameter
space, including two-loop gauge coupling unification \LNPZ\ yields a restricted
region of parameter space to be subjected to further phenomenological tests.

The no-scale flipped $SU(5)$ supergravity model \LNZb\ assumes that the
parameters $m_0$ and $A$ vanish, as is typically the case in no-scale
supergravity models \LN. This constraint reduces the dimension of the parameter
space down to three, and we take $m_t=100,130,160\GeV$. The choice of gauge
group (flipped $SU(5)$) is made to make contact with string-inspired models
which unify at the scale $M_U\sim10^{18}\GeV$. We also consider a ``strict
no-scale" scenario (see Sec. 4 of Ref. \LNZb) where in addition the universal
bilinear soft-supersymmetry breaking scalar mass parameter $B$ vanishes. This
constraint reduces the dimension of the parameter space down to two, since
$\tan\beta$ can now be computed as a function of $m_{1/2}$ and $m_t$. An
interesting consequence of this scenario is that for $\mu>0$ only
$m_t\lsim135\GeV$ is allowed, whereas for $\mu<0$ only $m_t\gsim140\GeV$ is
allowed. Moreover, for $\mu>0$ the calculated value of $\tan\beta$ can be
double-valued. In what follows we take $m_t=100,130\,(140,150,160)\GeV$ for
$\mu>0\,(\mu<0)$.

In both models there are semi-quantitative relations among the lighter
neutralino and chargino masses ($m_{\chi^0_1}\sim{1\over2}m_{\chi^0_2}
\sim{1\over2}m_{\chi^\pm_1}$) and it is found that these can be as light
as presently experimentally allowed. The gluino mass is bounded below
by $m_{\tilde g}\gsim200\GeV$ in both models, with $m_{\tilde q}\approx
m_{\tilde g}$ in the flipped model whereas $m_{\tilde q}>2m_{\tilde g}$
in the minimal $SU(5)$ model. Also, light (heavy) sleptons are characteristic
of the flipped (minimal) $SU(5)$ model. The Higgs spectrum of the minimal
model is to a good approximation SM-like \LNPWZh, in that the lightest Higgs
boson has close to SM couplings to gauge bosons and fermions, whereas the other
Higgs bosons ($A,H,H^\pm$) are heavy. In the flipped model something similar
occurs for a good portion of the allowed parameter space. However, significant
deviations can occur \LNPWZh\ which allow, \eg, charged Higgs masses as low as
$\approx100\GeV$.  In what follows we evaluate $\brbsg$ using as inputs the
couplings and masses obtained by running over all points in the previously
determined allowed parameters spaces for these models.

In Fig. 1 (top row) we present $\brbsg$ for the minimal $SU(5)$ supergravity
model, plotted against one of the light sparticle masses in the model, namely
the lightest chargino mass. We find
\eqn\B{2.3\,(2.6)\times10^{-4}<\brbsg_{minimal}<3.6\,(3.3)\times10^{-4},}
for $\mu>0\,(\mu<0)$, which are all within the new CLEO bound. Since these
numbers are within the SM range, in Fig. 1 (bottom row) we show their ratios to
the corresponding SM contribution, which depends solely on $m_t$. We obtain
\eqn\C{0.90\,(0.97)<\brbsg_{minimal}/\brbsg_{SM}<1.20\,(1.13),}
for $\mu>0\,(\mu<0)$. This implies that $\brbsg$ would need to be measured
with better than $20\%$ accuracy to start disentangling the minimal $SU(5)$
supergravity model from the SM. Moreover, for $\mu<0$ there is a band of
points which will be difficult to tell apart (requires $<1\%$ accuracy).
It is interesting to remark that an analogous set of plots versus $m_t$
instead, does not reveal any particular structure in $m_t$, thus an independent
determination of $m_t$ is not likely to constrain the possible values of
$\brbsg$. This is in sharp contrast with the SM, where such a measurement would
determine $\brbsg$, up to QCD uncertainties. In other words, in the minimal
$SU(5)$ supergravity model, a measurement of $\brbsg$ within the range of
Eq. (3), is unlikely to shed much light on the value of $m_t$. Of course, if
the measurement falls outside this range, the model would be excluded.

In Fig. 2 we present the analogous results for the no-scale flipped $SU(5)$
supergravity model (general case; strict no-scale scenario discussed below).
(In this figure $m_t=100,130,160\GeV$.) The results are strikingly different
than in the prior case.  One observes that part of the parameter space is
actually excluded by the new CLEO bound, for a range of sparticle masses.
However, light charged Higgs masses are not excluded altogether. Perhaps the
most surprising feature of the results is the strong suppression of $\brbsg$
which occurs for a good portion of the parameter space for $\mu>0$. Note also
that the results are generally quite different from the corresponding SM ones
(see Fig. 2, bottom row) and not much accuracy would be needed to tell the
models apart. Moreover, at least half of the parameter space should be
accessible with an increase in sensitivity by a factor of two. The dependence
on $m_t$ is also quite significant in this model, and for $m_t=130\GeV$ and
$\mu<0$, the calculated $\brbsg$ values are quite large and are starting to be
constrained: note the set or nearly parallel lines intersecting the CLEO bound
for $\mu<0$ (top row), as well as the corresponding isolated branch (bottom
row) for large values of the ratio.

In Fig. 3 we present the corresponding results for the strict no-scale scenario
in the flipped $SU(5)$ model. The results fall within those in Fig. 2 as
expected, but are a lot more predictive. For $\mu>0$ the two $\tan\beta$
solutions corresponding to $m_t=130\GeV$ give quite distinguishable results.
For $\mu<0$ the results for $m_t=140\GeV$ overlap with those for $m_t=150\GeV$.
Clearly, independent determinations of $m_t$ and $\brbsg$ would lead to very
precise constraints, as follows: (i) if $m_t\gsim140\GeV$ then $\mu<0$ and
$\brbsg\gsim3.5\times10^{-4}$; and (ii) if $m_t\lsim135\GeV$ then $\mu>0$ and
a wide range of $\brbsg$ values are possible. In both cases, with sufficiently
accurate measurements, one should be able to pin down the value of the chargino
mass (with a possible two-fold ambiguity) and therefore the whole spectrum of
the model.

The large suppression for $\mu>0$ in the flipped model (see Figs. 2,3) requires
further explanation. What happens is that in Eq. \A, the $A_\gamma$ term nearly
cancels against the QCD correction factor $C$; the $A_g$ contribution is small.
The $A_\gamma$ amplitude receives three contributions: from the $W$-$t$ loop
(\ie, the SM contribution), from the $H^\pm$-$t$ loop, and from the
$\chi^\pm_{1,2}$-$\tilde q$ loop. The first two contributions are
always negative \GSW, whereas the last one can have either sign. When the
chargino-squark contribution is positive and in the range $0.4-0.6$, the
cancellation occurs. No cancellation occurs for $\mu<0$ since there the
chargino-squark contribution is always below $\approx0.15$. Moreover, when this
contribution is positive and large ($\gsim1.2$ for $\mu>0$) or sufficiently
negative ($\lsim-0.05$ for $\mu<0$), the CLEO bound is exceeded. It has proven
quite difficult to determine a simple reason for the behavior of the
chargino-squark contribution and its sign-of-$\mu$ dependence. All we can do is
to remark that the lightest stop eigenstate $\tilde t_1$ is lighter for $\mu>0$
and it can approach the $\chi^\pm_2$ mass, resulting in an enhancement in
$A_\gamma$. A more detailed study of the circumstances under which the observed
phenomenon occurs would be better motivated should the experimentally
interesting range of $\brbsg$ fall below $\sim10^{-4}$. In fact, an actual
observation of the exclusive $B\to K^*\gamma$ mode has also been recently
reported (${\rm BR}(B\to K^*\gamma)=(4.5\pm1.5\pm0.9)\times10^{-5}$ \newCLEO),
which indicates that $\brbsg$ is probably not below $\sim10^{-5}$ and thus the
points in Figs. 2,3 with very suppressed $\brbsg$ values are probably
disfavored.

We conclude that the $\bsg$ microscope can probe the parameter spaces of the
two models we have considered in ways not possible with present high-energy
collider experiments. A sharpening of previous phenomenological predictions
for processes at Fermilab \LNWZ, LEPII \LNPWZ, and HERA \hera\ in the flipped
$SU(5)$ model would be possible in the light of the constraints imposed on the
parameter space of this model by the new CLEO bound \nino. With further
improvements in sensitivity to the inclusive $\bsg$ mode, it should not be long
before these models begin to show their true colors.

\bigskip
\bigskip
\bigskip
\bigskip
\noindent{\it Acknowledgments}: This work has been supported in part by DOE
grant DE-FG05-91-ER-40633. The work of J.L. has been supported by an SSC
Fellowship. The work of  D.V.N. has been supported in part by a grant from
Conoco Inc. The work of G.P. has been supported by a World-Laboratory
Fellowship. We would like to thank Kajia Yuan for spurring our interest on
this subject and his participation at the early stages of this project.
%\listrefsd
%\listrefs
\footatend\bigskip\bigskip\bigskip\immediate\closeout\rfile\writestoppt
\baselineskip=14pt\centerline{{\bf References}}\bigskip{\frenchspacing%
\parindent=20pt\escapechar=` \input refs.tmp\vfill\eject}\nonfrenchspacing
\listfigs
%\listfigsd
\bye